\documentclass[aps,floats]{revtex4}
\usepackage{amsmath,amssymb}
\usepackage{graphicx,epsfig}

\begin{document}
\bibliographystyle {plain}

\def\oppropto{\mathop{\propto}} 
\def\opsimeq{\mathop{\simeq}}
\def\opoverderline{\mathop{\overline}}
\def\operarrow{\mathop{\longrightarrow}}
\def\opsim{\mathop{\sim}} 
\def\opmin{\mathop{\min}} 
\def\opmax{\mathop{\max}} 

\def\fig#1#2{\includegraphics[height=#1]{#2}}
\def\figx#1#2{\includegraphics[width=#1]{#2}}


\title{ Anderson localization transition with long-ranged hoppings : \\
analysis of the strong multifractality regime in terms of weighted L\'evy sums } 


 \author{ C\'ecile Monthus and Thomas Garel }
  \affiliation{ Institut de Physique Th\'{e}orique, CNRS and CEA Saclay,
 91191 Gif-sur-Yvette, France}

\begin{abstract}
For Anderson tight-binding models in dimension $d$ with random on-site energies $\epsilon_{\vec r}$ and critical long-ranged hoppings decaying typically as $V^{typ}(r) \sim V/r^d$, we show that the strong multifractality regime corresponding to small $V$ can be studied via the standard perturbation theory for eigenvectors in quantum mechanics. The Inverse Participation Ratios $Y_q(L)$, which are the order parameters of Anderson transitions, can be written in terms of weighted L\'evy sums of broadly distributed variables (as a consequence of the presence of on-site random energies in the denominators of the perturbation theory). We compute at leading order the typical and disorder-averaged multifractal spectra $\tau_{typ}(q)$ and $\tau_{av}(q)$ as a function of $q$. For $q<1/2$, we obtain the non-vanishing limiting spectrum $\tau_{typ}(q)=\tau_{av}(q)=d(2q-1)$ as $V \to 0^+$. For $q>1/2$, this method yields the same disorder-averaged spectrum $\tau_{av}(q)$ of order $O(V)$ as obtained previously via the Levitov renormalization method by Mirlin and Evers [Phys. Rev. B 62,  7920 (2000)]. In addition, it allows to compute explicitly the typical spectrum, also of order $O(V)$, but with a different $q$-dependence $\tau_{typ}(q) \ne \tau_{av}(q)$ for all $q>q_c=1/2$. As a consequence, we find that the corresponding singularity spectra $f_{typ}(\alpha)$ and $f_{av}(\alpha)$ differ even in the positive region $f>0$, and vanish at different values $\alpha_+^{typ} > \alpha_+^{av}$, in contrast to the standard picture. We also obtain that the saddle value $\alpha_{typ}(q)$ of the Legendre transform reaches the termination point $\alpha_+^{typ}$ where $f_{typ}(\alpha_+^{typ} )=0 $ only in the limit $q \to +\infty$.

\end{abstract}

\maketitle

\section{ Introduction} 

Since its discovery fifty years ago \cite{anderson},
Anderson localization has remained a very active field
of research (see the reviews \cite{thouless,souillard,bookpastur,
Kramer,janssenrevue,markos,mirlinrevue}).
The order parameters of Anderson transitions 
are the inverse participation ratios (I.P.R.) 
of eigenfunctions $\psi (\vec r)$ on a finite volume $L^d$
\begin{eqnarray}
Y_q(L)  \equiv 
\frac{ \int_{L^d} d^d { \vec r}  \vert \psi (\vec r) \vert^{2q} }
{ \left[ \int_{L^d} d^d { \vec r}  \vert \psi (\vec r) \vert^{2}  \right]^q }
\label{defipr}
\end{eqnarray}
(the denominator can be omitted if the eigenfunction has been normalized
with $\int_{L^d} d^d { \vec r}  \vert \psi (\vec r) \vert^{2}=1 $).
As $L \to +\infty$, these I.P.R. converge to finite values in the localized phase, and behave as $L^{d(1-q)}$ in the delocalized phase.
At criticality, the eigenfunctions become multifractal and
 the I.P.R. involve non-trivial exponents \cite{janssenrevue,mirlinrevue}.
It is actually useful to introduce both
the typical and the averaged exponents \cite{mirlin_evers}
\begin{eqnarray}
 Y_q^{typ}(L) \equiv e^{\overline{ \ln Y_q(L) }}   && \opsimeq_{L \to \infty}  L^{ - \tau_{typ}(q) } \nonumber \\
\overline{ Y_q(L) }  && \opsimeq_{L \to \infty}  L^{ - \tau_{av}(q) }
\label{typetav}
\end{eqnarray}
The distribution $P_q(y_q)$
of the rescaled variable $y_q=Y_q(L)/Y_q^{typ}(L)$ is expected to decay as a power-law \cite{mirlin_evers,us_travellingrg}
\begin{eqnarray}
P_q(y_q) \oppropto_{y_q \to \infty} \frac{1}{y_q^{1+\beta_q}}
\label{defbetaq}
\end{eqnarray}
so that the typical and averaged exponents coincide or not
according to the value of $\beta_q$  \cite{mirlin_evers,mirlinrevue}
\begin{eqnarray}
\tau_{typ}(q) && =\tau_{av}(q) \ \ {\rm if} \ \ \beta_q>1 \nonumber \\
\tau_{typ}(q) && \ne \tau_{av}(q) \ \ {\rm if} \ \ \beta_q<1
\label{criterionbetaq}
\end{eqnarray}
In particular in the region $q>0$, one expects that there exists a critical value $q_c$ where $\beta_{q_c}=1$ separating the region where 
the two coincide for $q<q_c$ from the region where the two differ
with  \cite{mirlin_evers,mirlinrevue}
\begin{eqnarray}
 \beta_{q>q_c} = \frac{q_c}{q}
\label{betaqafterqc}
\end{eqnarray}

For the usual short-ranged Anderson tight-binding model
in finite dimension $d$, one expects a continuous interpolation 
\cite{mildenberger2002,mirlinrevue}
between a 'weak multifractality' regime obtained in
 the $d=2+\epsilon$ expansion \cite{Wegner_epsilon}
(the leading order corresponds to the Gaussian parabolic approximation 
of the multifractal spectrum) and a 'strong multifractality' (SM)
that occur in high dimension, with the following limiting form as $d \to +\infty$
\begin{eqnarray}
\tau^{SM}_{typ}(q)= \tau^{SM}_{av}(q) && = d (2q-1) \ \ {\rm for } \ \ q<\frac{1}{2} \nonumber \\
\tau^{SM}_{typ}(q)=\tau^{SM}_{av}(q) && = 0 \ \ {\rm for } \ \ q>\frac{1}{2}
\label{sm}
\end{eqnarray}

For Anderson tight-binding models
with long-ranged hoppings with typical asymptotic decay
\begin{eqnarray}
V^{typ}(r) \oppropto_{r \to +\infty}  \frac{ V}{r^a} 
\label{prbm}
\end{eqnarray}
one has also found a continuous family of the critical points  at $a=d$ 
as a function of the amplitude $V$ \cite{mirlin96}, that interpolates 
 between 'weak multifractality' for large $V \to +\infty$ 
and 'strong multifractality' for small $V \to 0$ \cite{mirlinrevue}.
In dimension $d=1$, corresponding to
 the ensemble of $L \times L$ power-law random banded matrices (PRBM),
various properties of these Anderson transitions have been studied in detail
\cite{mirlin_evers,varga00,kra06,garcia06,cuevas01,cuevas01bis,varga02,cuevas03,mildenberger,mendez05,mendez06,us_transmission}.   
In particular, the regime $ V \to 0$ has been analyzed via the powerful
Levitov renormalization method \cite{levitov,mirlin_evers},
that allows to compute exactly
the disorder-averaged spectrum in the region $q>1/2$ \cite{mirlin_evers}
\begin{eqnarray}
\tau^{PRBM}_{av}(q)= \frac{4b}{\sqrt \pi} 
\frac{ \Gamma \left( q-\frac{1}{2} \right)}{\Gamma(q-1)}
\label{taulevitovrg}
\end{eqnarray}
where $b=V/W$ is the ratio between the amplitude $V$
and the width $W$ of the on-site random energies.
The limiting 'strong multifractality' spectrum of Eq. \ref{sm}
has been then obtained in \cite{mirlin_evers} by using
the symmetry of the multifractal spectrum 
\cite{mirlin06,milden07,qHall,vasquez,us_symfalpha}
connecting exponents with $q<1/2$ to exponents with $q>1/2$
\begin{eqnarray}
\tau_{av}(q)-\tau_{av}(1-q)=d(2q-1)
\label{symtauq}
\end{eqnarray}
which is expected to hold for any Anderson transition in the so-called 
'conventional symmetry classes' \cite{mirlinrevue}.
The derivation of Eq. \ref{sm} is thus rather indirect since
one performs an explicit calculation in the region $q>1/2$
to obtain that the exponents vanish $\tau_{av}(q>1/2) \to 0$
in this region, and the non-vanishing results in the region $q<1/2$
are then entirely based on the symmetry of Eq. \ref{symtauq}.
The same methodology has been followed for related hierarchical models
\cite{fyodorov} and for the analysis of short-ranged models in high $d$
\cite{mildenberger2002}.

The aim of this paper is to reconsider the regime of 'strong multifractality' $V \to 0$ for Anderson tight-binding models with long-ranged hoppings 
 in order to compute explicitly both the typical and the
disorder-averaged multifractal spectra in both regimes $0<q<1/2$ and $q>1/2$.
The paper is organized as follows.
In section \ref{sec_perturbation}, we describe the model and apply the standard
perturbation theory of quantum mechanics to obtain the leading order
expression of the I.P.R. $Y_q$ of Eq. \ref{defipr}.
In section \ref{levysum}, we analyze the statistical properties of some weighed L\'evy sums that play a major role in this perturbation theory.
The typical behaviors and the disorder-averaged behaviors of the I.P.R. $Y_q$
are then computed in sections 
\ref{typical} and \ref{averaged} respectively.
Our final results concerning the typical and averaged multifractal spectra
in the various regions of $q$  are discussed in section \ref{discussion} and 
are compared to previous results.
Our conclusions are summarized in section \ref{conclusion}.

\section{ Perturbation theory for Anderson tight-binding models with long-ranged hoppings }

\label{sec_perturbation}

\subsection{ Anderson tight-binding models with long-ranged hoppings }

We consider an Anderson tight-binding model 
on an hypercubic lattice of size $L^d$ defined by the Hamiltonian 
\begin{eqnarray}
H && =H_0+H_1 \nonumber \\
H_0 && = \sum_{\vec r} \epsilon_{\vec r} \vert \vec r > < \vec r \vert  \nonumber \\
H_1 && = \sum_{\vec r \ne \vec r \ '} 
V_{\vec r,\vec r \ '}  \vert \vec r > <  \vec r \ '\vert 
\label{Hperturbation}
\end{eqnarray}
where the hoppings are symmetric $V_{\vec r \ ',\vec r }=V_{\vec r,\vec r \ '}$
with $V_{\vec r,\vec r } =0$.

\subsubsection{ Assumptions on the random on-site energies $\epsilon_{\vec r} $}

We consider the usual case where the on-site energies $\epsilon_{\vec r}$
are independent random variables distributed with a  
law $P(\epsilon)$ which is symmetric around $\epsilon=0$
and which presents the scaling form
\begin{eqnarray}
P(\epsilon)= \frac{1}{W} p \left( \frac{\epsilon}{W} \right)
  \label{pepsW}
\end{eqnarray} 
so that $W$ represents the disorder strength.
In the following, an important role will be played by
the probability density around $\epsilon=0$ that will be denoted by
\begin{eqnarray}
P(\epsilon=0)= \frac{c}{ 2 W} 
  \label{pepsWzero}
\end{eqnarray} 
where $c=2 p \left( 0 \right)>0$.
For instance in the PRBM model \cite{mirlinrevue}, $P(\epsilon)$ is
a Gaussian of variance unity corresponding to the values
\begin{eqnarray}
W^{PRBM} && =1 \nonumber \\
c^{PRBM} && =\sqrt{\frac{2}{\pi}}
  \label{epsPRBM}
\end{eqnarray} 

The average over the random energies $\{\epsilon_{\vec r}\}$
will be denoted by 
\begin{eqnarray}
E \left( A(\{\epsilon_{\vec r}\})  \right) \equiv 
\left[ \prod_{\vec r } \int d\epsilon_{\vec r} P(\epsilon_{\vec r})
\right] 
 A(\{\epsilon_{\vec r}\})
  \label{aveps}
\end{eqnarray}

\subsubsection{ Assumptions on the long-ranged hoppings
 $V_{\vec r,\vec r \ '}$}

We consider the critical case where the long-ranged hoppings decay
typically as $1/r^d$ in dimension $d$ \cite{anderson,mirlin96}
\begin{eqnarray}
V(\vec r) \opsimeq_{r \to \infty}
 \frac{V}{r^d} u_{\vec r}
\label{longranged}
\end{eqnarray}
where $u_{\vec r}$ are either fixed ($u_{\vec r} =1$)
or are independent random variable of order $O(1)$.
For instance in the PRBM model in $d=1$ \cite{mirlinrevue}, $u_{\vec r}$ is
a Gaussian of variance unity. In any case, we assume here that the distribution
of $u_{\vec r}$ is such that its moments exist.

In the following, two quantities will play a major role.
Denoting by $S_d$ the surface appearing in the radial change of variables
$d^d \vec r = S_d r^{d-1} dr$ in dimension $d$, we may evaluate
the behavior in $L$ of the following sums
\begin{eqnarray}
\sum_{\vec r  }  \overline{ \vert  V( \vec r ) \vert }
= S_d V \overline{ \vert u_{\vec r} \vert} \ln L
\label{sumvabs}
\end{eqnarray}
and for $q<1/2$
\begin{eqnarray}
\sum_{\vec r  }  \overline{ \vert  V( \vec r ) \vert }^{2q}
= V^{2q} \overline{\vert u_{\vec r} \vert^{2q}}  
\int S_d r^{d-1} dr     \frac{1}{r^{d2q}} 
  =  V^{2q} \overline{ \vert u_{\vec r} \vert^{2q}} S_d
\frac{ L^{d(1-2q)}  }{ d(1-2q)}
\label{sumv2q}
\end{eqnarray}

\subsection{ Perturbation theory in the hoppings  }

As explained in the Introduction, we focus in this paper on the
'strong multifractality regime' that corresponds to a small amplitude $V$
in the long-ranged hoppings of Eq. \ref{longranged}.
It is thus natural to consider the perturbation theory
associated to the decomposition of Eq. \ref{Hperturbation}.
The Hamiltonian $H_0$ has for eigenstates the $L^d$ completely localized
eigenfunctions on each lattice site, the eigenvalues being simply the
corresponding random on-site energies $\epsilon_{\vec r}$
\begin{eqnarray}
\vert \phi_{\vec r}^{(0)} > && = \vert \vec r > \nonumber \\
E_{\vec r}^{(0)} && = \epsilon_{\vec r}
\label{phi0}
\end{eqnarray}

 The standard perturbation theory of quantum mechanics yields that at lowest order,
the eigenvalues are unchanged
\begin{eqnarray}
E_{\vec r}^{(1)}  = \epsilon_{\vec r}
\label{ener1}
\end{eqnarray}
whereas the eigenfunctions read
\begin{eqnarray}
\vert \phi_{\vec r}^{(1)} > && = \vert \phi_{\vec r}^{(0)} > + 
\sum_{\vec r \ ' \ne \vec r } 
\vert \phi_{\vec r \ '}^{(0)} >
\frac{ < \phi_{\vec r \ '}^{(0)} \vert H_1 \vert \phi_{\vec r}^{(0)} > }{E_{\vec r}^{(0)}- E_{\vec r \ '}^{(0)}} 
  = \vert \vec r > + \sum_{\vec r \ ' \ne \vec r} \vert \vec r \ ' > 
\frac{ V_{\vec r, \vec r \ '}  }{\epsilon_{\vec r}-\epsilon_{\vec r \ '}  } 
\label{phi1}
\end{eqnarray}

The corresponding I.P.R. of Eq. \ref{defipr} read
\begin{eqnarray}
Y_q^{(1)}= \frac{ \sum_{\vec r \ '}
 \vert \phi _{\vec r}^{(1)}(\vec r \ ') \vert^{2q} }
{ \left[  \sum_{\vec r \ '} \vert \phi_{\vec r}^{(1)} (\vec r \ ') \vert^{2}  \right]^q }
= \frac{1+\Sigma_q}{ (1+\Sigma_1 )^q }
  \label{yq1self}
\end{eqnarray} 
in terms of the sums
\begin{eqnarray}
\Sigma_q  \equiv  \sum_{\vec r \ ' \ne \vec r}  \left\vert
 \frac{ V_{\vec r,\vec r \ '}  }{\epsilon_{\vec r}-\epsilon_{\vec r \ '}  }\right\vert^{2q}
  \label{defsigmaqfull}
\end{eqnarray} 
To simplify the notations from now on, we will focus 
on the eigenstate associated to the
central site $\vec r=0$ and we will consider that its associated 
eigenvalue is exactly at the center of the band 
 $\epsilon_{\vec 0}=0$. Then the perturbed eigenenergy of Eq. \ref{ener1}
is also $E_{\vec 0}^{(1)}  = \epsilon_{\vec 0}=0$ at leading order,
and the I.P.R. of the corresponding eigenstate (Eq. \ref{yq1self})
will then allows to compute at leading order
the multifractal spectrum corresponding to $E=0$.
The variables $\Sigma_q$ of Eq. \ref{defsigmaqfull} become
\begin{eqnarray}
\Sigma_q  \equiv  \sum_{\vec r \ne \vec 0}  \left\vert \frac{ V(\vec r)  }{\epsilon_{\vec r}  }\right\vert^{2q}
  \label{defsigmaq}
\end{eqnarray} 

The aim of this paper is to analyze the statistical properties
of the I.P.R. of Eq. \ref{yq1self}, in particular their typical values
and their disorder-averaged values to extract the multifractal exponents
$\tau_{typ}(q)$ and $\tau_{av}(q)$ of Eq. \ref{typetav}.
It is convenient to analyze first the statistics of the sums $\Sigma_q$
that turned out to be weighted L\'evy sums as we now explain.

\section{ Statistics of the weighted L\'evy sums $\Sigma_q$ }

\label{levysum}

In this section, we discuss the statistical properties of the sums $\Sigma_q$
of Eq. \ref{defsigmaq} that can be rewritten as sums
\begin{eqnarray}
\Sigma_q \equiv 
\sum_{\vec r \in L^d }  \vert  V( \vec r ) \vert^{2q} z_q(\vec r)
\label{sigmaq}
\end{eqnarray}
of the random variables
\begin{eqnarray}
z_q(\vec r) \equiv \vert \epsilon_{\vec r } \vert^{-2 q}
\label{zq}
\end{eqnarray}
with the weights $\vert  V( \vec r ) \vert^{2q}$.

We recall here that the average of an observable $\cal O$
over the random on-site energies will be denoted
by $E(\cal O)$ (see Eq. \ref{aveps}).
In the case where the long-ranged hoppings of Eq. \ref{longranged}
are non-random ($u_{\vec r} \equiv 1$), the disorder-average denoted by 
$\overline {\cal O}$ is equal to $E(\cal O)$.
In the case where the long-ranged hoppings of Eq. \ref{longranged}
are also random, the disorder-average denoted by 
$\overline {\cal O}$ denotes the average over both the random on-site energies
and the random variables $u_{\vec r}$ appearing in
the long-ranged hoppings of Eq. \ref{longranged}.

\subsection{ Statistics of the variables $z_q(\vec r) \equiv \vert \epsilon_{\vec r } \vert^{-2 q}$ }

From the probability density $P(\epsilon=0)$ near zero energy
given in Eq. \ref{pepsWzero}, one obtains via a change of variable 
that the probability distribution $Q_q(z_q)$ of the
variable $z_q(\vec r) \equiv \vert \epsilon_{\vec r } \vert^{-2 q}$
presents the following power-law decay
\begin{eqnarray}
Q_q(z_q) \opsimeq_{z_q \to +\infty} \frac{c \mu_q }{  W z_q^{1+\mu_q}}
\ \ {\rm with } \ \ \mu_q=\frac{1}{2q}
\label{lawzq}
\end{eqnarray}

In particular, the disorder-averaged value $E(z_q)$ (with the notation
of Eq. \ref{aveps})
presents a transition at $q=1/2$
\begin{eqnarray}
E(z_q) && < +\infty \ \ {\rm for } \ \ q > \frac{1}{2} \nonumber \\
E(z_q) && = +\infty \ \ {\rm for } \ \ q \leq \frac{1}{2}
\label{avzq}
\end{eqnarray}

\subsection{ Generating function $E (e^{-t \Sigma_q})$ of the sum $\Sigma_q$ }

L\'evy sums of identically broadly distributed variables
(without the weights $\vert  V( \vec r ) \vert^{2q}$ in Eq. \ref{sigmaq})
appears in various fields of disordered systems, 
usually in low-temperature disorder-dominated phases of classical models :
their statistical properties are described in particular in
\cite{jpbreview,Der,Der_Fly,us_critiweights}.
In the following, we analyze the effects of the presence of the weights 
$\vert  V( \vec r ) \vert^{2q}$. 

The first important property is that the disorder-averaged value
$E \left( \Sigma_q  \right) $ presents the same phase transition as Eq.
\ref{avzq}, as a consequence of the linearity of the sum of Eq. \ref{sigmaq}.

\subsubsection{ Case $q<1/2$ where the disorder-averaged value 
$E \left( \Sigma_q  \right)$ is finite }

For $q<1/2$, the disorder-averaged value is finite and reads
\begin{eqnarray}
E \left( \Sigma_q  \right) =
 E \left( z_q \right) \sum_{\vec r \in L^d }  \vert  V( \vec r ) \vert^{2q}
\label{moyqsmallE}
\end{eqnarray}
where using Eq. \ref{pepsW}
\begin{eqnarray}
 E \left( z_q \right) = \int_{-\infty}^{+\infty} d\epsilon P(\epsilon) \vert \epsilon \vert^{-2q}  
 = W^{-2q} B_q \ \ { \rm with } \ \ B_q =2 \int_{0}^{+\infty} dx p(x) x^{-2q}
\label{zqav}
\end{eqnarray}
For instance if $p(x)$ is Gaussian of variance unity as in the PRBM model,
one obtains
\begin{eqnarray}
 B_q^{Gauss}  =2 \int_{0}^{+\infty} dx \frac{e^{-\frac{x^2}{2}} }{\sqrt{2 \pi}} x^{-2q}
= \frac{ \Gamma \left( \frac{1}{2}-q \right)}{2^q \sqrt{ \pi}}
\label{bq}
\end{eqnarray}

After averaging also over the possibly random hoppings of Eq. \ref{longranged}, 
we finally obtain using Eq. \ref{sumv2q}
\begin{eqnarray}
\overline{ \Sigma_q  } = 
   \left( \frac{V}{W} \right)^{2q} B_q
S_d  \overline{ \vert u_{\vec r} \vert^{2q}} 
\frac{ L^{d(1-2q)}  }{d(1-2q)} 
\label{moyqsmall}
\end{eqnarray}

\subsubsection{ Case $q>1/2$ where the disorder-averaged value 
$E \left( \Sigma_q  \right)$ is infinite}

For $q>1/2$ where $\mu_q<1$, the generating function $E \left(e^{-t z_q} \right)$
presents the characteristic singularity in $t^{\mu_q}$ 
of L\'evy distribution (see Eq. \ref{lawzq})
\begin{eqnarray}
E \left(e^{-t z_q} \right)  && \equiv \int_0^{+\infty} dz_q Q_q(z_q) e^{-t z_q}
= 1 - \int_0^{+\infty} dz_q Q_q(z_q) (1-e^{-t z_q})
\opsimeq_{t \to 0} 1- t^{\mu_q} 
\frac{c\mu_q }{ W} \left[ - \Gamma(- \mu_q) \right] +... \nonumber \\
&& \opsimeq_{t \to 0} e^{ - t^{\mu_q} 
\frac{c \mu_q }{ W} \left[ - \Gamma(- \mu_q) \right] +...}
\label{lawzqlaplace}
\end{eqnarray}
with the usual integral
\begin{eqnarray}
 \int_0^{+\infty} \frac{dx}{x^{1+\mu_q}} (1-e^{- x}) = - \Gamma(- \mu_q) 
\label{integlevy}
\end{eqnarray}

The generating function $E \left( e^{-t \Sigma_q } \right)$
 will thus presents a similar singularity
\begin{eqnarray}
E \left( e^{-t \Sigma_q } \right) && = 
 \prod_{\vec r \in L^d } 
E \left( e^{-t   \vert V(\vec r) \vert^{2q} z_q } \right) \opsimeq_{t \to 0}
 e^{ - \sum_{\vec r} \left( t  \vert V(\vec r) \vert^{2q} \right)^{\mu_q} 
\frac{c \mu_q}{ W} \left[ - \Gamma(- \mu_q) \right] }
 \nonumber \\
&& \opsimeq_{t \to 0}
 e^{ - t^{\mu_q} \frac{c \mu_q }{ W} \left[ - \Gamma(- \mu_q) \right]
\sum_{\vec r}   \vert V(\vec r) \vert
 }
\label{sigmaqlaplace}
\end{eqnarray}

After averaging also over the possibly random hoppings of Eq. \ref{longranged}, 
we finally obtain using Eq. \ref{sumvabs}
\begin{eqnarray}
\overline{ e^{-t \Sigma_q } } && 
  \opsimeq_{t \to 0}
 1 - t^{\mu_q} \frac{c \mu_q }{ W} \left[ - \Gamma(- \mu_q) \right]
\sum_{\vec r}    \overline{\vert V(\vec r) \vert } 
 \nonumber \\
&& \opsimeq_{t \to 0}
 e^{- t^{\mu_q} \frac{c \mu_q V}{ W} \left[ - \Gamma(- \mu_q) \right]
\overline{\vert u_{\vec r} \vert } 
S_d \ln L  }  
\label{sigmaqlaplaceres}
\end{eqnarray}

Equivalently, inverting Eq. \ref{lawzqlaplace}, one obtains the 
following asymptotic decay for the probability distribution ${\cal P}_q (\Sigma_q)$
\begin{eqnarray}
{\cal P}_q (\Sigma_q)  \opsimeq_{\Sigma_q \to +\infty}
 \frac{ {\cal A}_q }{  \Sigma_q^{1+\mu_q}}
\label{lawsigmaq}
\end{eqnarray}
with the exponent $\mu_q=1/(2q)$ and the amplitude
\begin{eqnarray}
{\cal A}_q \equiv 
c S_d \mu_q \frac{V}{W}  \overline{\vert u_{\vec r} \vert}  \ln L  
\label{ampliq}
\end{eqnarray}

\subsection{ Analysis of the auxiliary quantity 
$E\left( z_1^q e^{-t z_1} \right)$ }

The variables $\Sigma_q$ and $\Sigma_1$ that appear in the numerator
and denominator of Eq. \ref{yq1self} are correlated since they involve 
the same random energies. In the following, some computations
will involve the auxiliary quantity
\begin{eqnarray}
E\left( z_1^q e^{-t z_1} \right)
\label{auxi}
\end{eqnarray}

From Eq. \ref{lawzq}, the probability density $Q_1(z_1)$ 
is know to decay with a power-law of exponent $1+\mu_1=3/2$
\begin{eqnarray}
Q_1(z_1) \opsimeq_{z_1 \to +\infty} \frac{c}{ 2 W z_1^{3/2}}
\label{lawzq1}
\end{eqnarray}

As a consequence for $q<1/2$, the non-integer
moment $E(z_1^q)$ of order $q$ exists and one has
\begin{eqnarray}
E\left( z_1^q e^{-t z_1} \right)  \opsimeq_{t \to 0} E\left( z_1^q  \right) 
+o(t)
\label{auxiqsmall}
\end{eqnarray}
For $q>1/2$, the non-integer
moment $E(z_1^q)$ of order $q$ does not exist, and the divergence of the auxiliary
quantity as $t \to 0$ can be obtained via the change of variables $x=t z_1$
\begin{eqnarray}
E\left( z_1^q e^{-t z_1} \right) && \equiv
\int_0^{+\infty} dz_1 Q_1(z_1) z_1^q e^{-t z_1}
= \int_0^{+\infty} \frac{dx}{t} Q_1 \left( \frac{x}{t} \right) 
\left( \frac{x}{t} \right)^q e^{- x} \nonumber \\
&& \opsimeq_{t \to 0}  \frac{c}{ 2 W } t^{\frac{1}{2}-q}
\int_0^{+\infty} dx^{\frac{3}{2}-q} x e^{- x}
=  \frac{c}{ 2 W } t^{\frac{1}{2}-q} \Gamma \left( q-\frac{1}{2}\right)
\label{auxiqlarge}
\end{eqnarray}

\section{ Typical values of the I.P.R. $Y_q$ }

\label{typical}

To obtain the typical values of the I.P.R. $Y_q$ (see Eq. \ref{typetav}),
we need to average the logarithm of the expression of Eq. \ref{yq1self}
\begin{eqnarray}
\ln Y_q^{typ} \equiv \overline{ \ln Y_q }   = \overline{ \ln \left(1+ \Sigma_q \right) }
- q \overline{ \ln \left( 1+ \Sigma_1 \right) }
\label{Yqtyp}
\end{eqnarray}
So here the correlations between $\Sigma_q$ and $\Sigma_1$ do not play any role,
and one only needs to know the statistical properties of the sums $\Sigma_q$
and $\Sigma_1$ discussed in the previous section.

\subsection{ Computation of $\overline{ \ln \left(1+ \Sigma_q \right) }$ for $q<1/2$ } 

For $q<1/2$, the disorder-averaged value $\overline{ \Sigma_q}$ converges
(see Eq. \ref{moyqsmall}). So we may use the expansion of the logarithm to 
obtain the leading-order behavior
\begin{eqnarray}
 \overline{ \ln  \left(1+ \Sigma_q \right) }
&& \simeq \overline{ \Sigma_q  } =
 \left( \frac{V}{W} \right)^{2q} B_q
S_d  \overline{ \vert u_{\vec r} \vert^{2q}} 
\frac{ L^{d(1-2q)}  }{d(1-2q)} 
\label{moylogsigmaqsmall}
\end{eqnarray}

\subsection{ Computation of $\overline{ \ln \left(1+ \Sigma_q \right) }$ for $q>1/2$ } 

For $q>1/2$, we use the following integral representation of the logarithm
\begin{eqnarray}
\ln (1+ \Sigma_q) = \int_0^{+\infty} \frac{dt}{t} e^{-t} \left( 1-e^{- t \Sigma_q} \right)
\label{logrepresentation}
\end{eqnarray}
to relate the disorder-averaged value to the generating function 
$\overline {e^{- t \Sigma_q} } $
\begin{eqnarray}
\overline{ \ln (1+ \Sigma_q) }= \int_0^{+\infty} \frac{dt}{t} e^{-t} 
\left( 1- \overline {e^{- t \Sigma_q} } \right)
\label{logcalculsigmaq}
\end{eqnarray}

Using the result of Eq. \ref{sigmaqlaplaceres}, we obtain
\begin{eqnarray}
\overline{ \ln (1+ \Sigma_q) } && = \int_0^{+\infty} \frac{dt}{t} e^{-t} 
\left( 1- \overline {e^{- t \Sigma_q} } \right)
\simeq \int_0^{+\infty} \frac{dt}{t} e^{-t} 
\left[  t^{\mu_q} \frac{c \mu_q V}{W} \left[ - \Gamma(- \mu_q) \right]
\overline{\vert u_{\vec r} \vert } 
S_d \ln L \right] \nonumber \\
&& \simeq S_d \ln L
\frac{c \mu_q V }{W} \overline{\vert u_{\vec r} \vert}
 \left[ - \Gamma(- \mu_q) \right] \Gamma(\mu_q) 
\label{logcalculsigmaqbis}
\end{eqnarray}
Using the relation
\begin{eqnarray}
 - \Gamma(- \mu_q) \Gamma(\mu_q)= \frac{ \pi }{\mu_q \sin(\pi \mu_q)  }
\label{gamma2}
\end{eqnarray}
the final result reads
\begin{eqnarray}
\overline{ \ln (1+ \Sigma_q) }  \simeq S_d \ln L
\frac{c V }{W} \overline{\vert u_{\vec r} \vert}
 \frac{ \pi }{ \sin(\pi \mu_q)  }
\label{reslogq}
\end{eqnarray}

Note that this result can be equivalently obtained via a direct calculation based
on the asymptotic behavior of Eq.  \ref{lawsigmaq}
\begin{eqnarray}
\overline{ \ln (1+ \Sigma_q) } &&  = \int_0^{+\infty} d\Sigma_q
{\cal P}_q (\Sigma_q)  \ln (1+ \Sigma_q)
\simeq {\cal A}_q \int_0^{+\infty} \frac{d\Sigma_q}{  \Sigma_q^{1+\mu_q}} \ln (1+ \Sigma_q)
= {\cal A}_q \frac{\pi}{\mu_q \sin(\pi \mu_q)} \nonumber \\
&& = S_d \ln L \overline{\vert u_{\vec r} \vert}  \frac{c  V  }{  W }
\frac{\pi}{ \sin(\pi \mu_q)}
\label{reslogqbis}
\end{eqnarray}

\subsection{ Behavior of the typical I.P.R. $Y_q$ for $0<q<1/2$ }

For $q<1/2$, we have seen that $\overline{ \ln 
\left(1+ \Sigma_q \right) }$ grows as $L^{d(1-2q)}$  
(cf \ref{moylogsigmaqsmall}), whereas $\overline{ \ln 
\left(1+ \Sigma_1 \right) }$ grows only as $\ln L$ (Eq \ref{reslogq}).
We thus obtain at leading order 
\begin{eqnarray}
 Y_q^{typ}(L) \equiv e^{ \overline{\ln Y_q} }
 \sim  e^{  \overline{\Sigma_q  } } \sim 
e^{  \left( \frac{V}{W} \right)^{2q}  S_d B_q \overline{ \vert u_{\vec r}  \vert^{2q} }
\frac{ L^{d(1-2q)} }{d(1-2q)} } \simeq  1+  \left( \frac{V}{W} \right)^{2q}
S_d B_q \overline{ \vert u_{\vec r} \vert^{2q} } 
\frac{ L^{d(1-2q)} }{d(1-2q)}
\label{resiprtypqsmall}
\end{eqnarray}
The typical exponents $\tau_{typ}(q)$ defined in Eq. \ref{typetav} thus read
\begin{eqnarray}
 \tau_{typ}(q<1/2) =  d  (2q-1) 
\label{restautypqsmallerdemi}
\end{eqnarray}
in agreement with the 'strong multifractality' limit of Eq. \ref{sm}.

\subsection{ Behavior of the typical I.P.R. $Y_q$ for $q>1/2$ } 

For $q>1/2$, we have seen that both $\overline{ \ln 
\left(1+ \Sigma_q \right) }$ and $\overline{ \ln 
\left(1+ \Sigma_1 \right) }$ grow as $\ln L$ (Eq. \ref{reslogq}).
These two contributions yield at leading order
\begin{eqnarray}
\overline{ \ln Y_q }   = \overline{ \ln 
\left(1+ \Sigma_q \right) }
- q \overline{ \ln 
\left( 1+ \Sigma_1 \right) }
= S_d \ln L
\frac{c V }{W} \overline{\vert u_{\vec r} \vert}
\left[ \frac{ \pi }{ \sin(\pi \mu_q)  } 
- q \frac{ \pi }{ \sin(\pi \mu_1) } \right]
\label{Yqtypreslevy}
\end{eqnarray}
Since $\mu_1=1/2$, we obtain that
the typical exponents $\tau_{typ}(q)$ defined in Eq. \ref{typetav} read
\begin{eqnarray}
\tau_{typ}(q>1/2) = S_d 
\frac{c V }{W} \overline{\vert u_{\vec r} \vert}  \pi 
\left[ q - \frac{ 1 }{ \sin ( \frac{\pi}{2q} )  }    \right]
\label{restauqtypreslevy}
\end{eqnarray}

\section{ Disorder-averaged values of the I.P.R. $Y_q$ }

\label{averaged}

To compute the disorder-averaged values of the I.P.R. $Y_q$ of Eq. \ref{yq1self},
it is convenient to use the identity
\begin{eqnarray}
\frac{1}{a^q} = \frac{1}{\Gamma(q)} \int_0^{+\infty} dt \  t^{q-1} e^{-at} 
\label{defgamma}
\end{eqnarray}
to obtain
\begin{eqnarray}
\overline{ Y_q}  && = \overline{ Y_q }\vert_{first \  contribution} 
+ \overline{ Y_q }\vert_{second \  contribution} \nonumber \\
 \overline{ Y_q }\vert_{first \  contribution} && =
\frac{1}{\Gamma(q)} \int_0^{+\infty} dt \ t^{q-1} e^{-t}
\ \overline{  e^{-t \Sigma_1 }}
\nonumber \\
 \overline{ Y_q }\vert_{second \  contribution} && =
\frac{1}{\Gamma(q)} \int_0^{+\infty} dt \ t^{q-1} e^{-t} \ 
\overline{ \Sigma_q   e^{-t \Sigma_1 }}
\label{Yqmoyperturbation}
\end{eqnarray}
We now evaluate separately these two contributions.

\subsection{ Computation of the first contribution }

Using Eq. \ref{sigmaqlaplaceres} for $q=1$ with $\mu_1=1/2$
and $\left[ - \Gamma(- 1/2) \right]=2 \sqrt{\pi}$
\begin{eqnarray}
\overline{ e^{-t \Sigma_1 } } \simeq
 e^{- t^{1/2} \frac{c V  \sqrt{\pi} }{W}
\overline{\vert u_{\vec r} \vert } S_d \ln L  }  
\label{sigmaqlaplaceresq1}
\end{eqnarray}
the first contribution of Eq. \ref{Yqmoyperturbation} reads at leading order
\begin{eqnarray}
\overline{ Y_q }\vert_{first contribution} 
&& \equiv \frac{1}{\Gamma(q)} \int_0^{+\infty} dt \ t^{q-1} e^{- t}
\overline{  e^{-t \Sigma_1 }}
 \simeq \frac{1}{\Gamma(q)} \int_0^{+\infty} dt \ t^{q-1} e^{- t}
\left(1 - t^{1/2} \frac{c V  \sqrt{\pi} }{W}
\overline{\vert u_{\vec r} \vert } S_d \ln L   \right) \nonumber \\
&& \simeq 1 - \frac{\Gamma \left( q+\frac{1}{2} \right)}{\Gamma(q)} \frac{c V  \sqrt{\pi} }{W}
\overline{\vert u_{\vec r} \vert } S_d \ln L
\label{firstcontri}
\end{eqnarray}

\subsection{ Computation of the second contribution }

To evaluate the second  contribution of Eq. \ref{Yqmoyperturbation},
we first need to evaluate, using the definitions of Eqs \ref{sigmaq} and \ref{zq}
\begin{eqnarray}
E\left( \Sigma_q e^{-t \Sigma_1} \right)
&& = E\left( \sum_{\vec r  }  \vert  V( \vec r ) \vert^{2q} 
 z_1^q(\vec r) 
e^{-t  \sum_{\vec r \ '  }  \vert  V( \vec r \ ' ) \vert^{2} 
 z_1(\vec r \ ') } \right) \nonumber \\
&& =  \sum_{\vec r  }  \vert  V( \vec r ) \vert^{2q}
 E\left(  z_1^q(\vec r) e^{ -t  \vert  V( \vec r  ) \vert^{2} 
 z_1(\vec r ) } \right)
E \left(  e^{-t  \sum_{\vec r \ ' \ne \vec r }  \vert  V( \vec r \ ' ) \vert^{2} 
 z_1(\vec r \ ') } \right)
\label{auxisum}
\end{eqnarray}

For $q<1/2$, we use Eqs \ref{auxiqsmall} and \ref{sigmaqlaplaceresq1}
to obtain
\begin{eqnarray}
E\left( \Sigma_q e^{-t \Sigma_1} \right)
 \opsimeq_{t \to 0} 
E\left( z_1^q  \right) \sum_{\vec r  }  \vert  V( \vec r ) \vert^{2q}
\label{auxisumqsmall}
\end{eqnarray}
so that the second contribution reads at leading order using Eq. \ref{sumv2q}
\begin{eqnarray}
 \overline{ Y_{q<1/2} }\vert_{second contribution} 
\simeq  \overline{ \Sigma_q }  = E\left( z_1^q  \right) \sum_{\vec r  }  \overline{
\vert  V( \vec r ) \vert^{2q} }
\simeq  E\left( z_1^q  \right)
S_d V^{2q} \overline{ \vert  u_{\vec r} \vert^{2q} } \frac{ L^{d(1-2q)} }{d(1-2q)}
\label{secondcontriqsmallerdemifin}
\end{eqnarray}

For $q>1/2$, we use Eqs  \ref{auxiqlarge} and \ref{sigmaqlaplaceresq1}
to evaluate the singularity of Eq. \ref{auxisum} for small $t$
\begin{eqnarray}
E\left( \Sigma_q e^{-t \Sigma_1} \right)
&& \opsimeq_{t \to 0}  \sum_{\vec r  }  \vert  V( \vec r ) \vert^{2q}
 \frac{c }{ 2 W } \left( t V^2( \vec r ) \right)^{\frac{1}{2}-q} \Gamma \left( q-\frac{1}{2}\right)
\left( 1 - t^{1/2} \frac{c V  \sqrt{\pi} }{W}
\overline{\vert u_{\vec r} \vert } S_d \ln L   +... \right) \nonumber \\
&& \opsimeq_{t \to 0} t^{\frac{1}{2}-q}
 \sum_{\vec r  }  \vert  V( \vec r ) \vert
 \frac{c}{ 2 W }  \Gamma \left( q-\frac{1}{2}\right)+...
\label{auxisumqlarge}
\end{eqnarray}
Using Eq. \ref{sumvabs}, we finally obtain at leading order
\begin{eqnarray}
\overline{ Y_{q>1/2} }\vert_{second contribution} 
 && = \frac{1}{\Gamma(q)} \int_0^{+\infty} dt \ t^{q-1} e^{-t}
\overline{ \Sigma_q   e^{-t \Sigma_1 }}
\simeq  \frac{1}{\Gamma(q)} 
 \frac{c}{2 W } \Gamma(1/2) \Gamma \left( q-\frac{1}{2}\right)
\sum_{\vec r  }  \overline{ \vert  V( \vec r ) \vert } \nonumber \\
&& \simeq \frac{\Gamma \left( q-\frac{1}{2}\right)}{\Gamma(q)} 
 \frac{c \sqrt{\pi}}{2 W }  
S_d V \overline{ \vert u_{\vec r} \vert} \ln L
\label{secondcontriqbiggererdemifin}
\end{eqnarray}

\subsection{ Disorder-averaged I.P.R. for $q<1/2$ }

For $q<1/2$, the first contribution of order $\ln L$ (Eq. \ref{firstcontri})
is negligible with respect to the second contribution
of Eq. \ref{secondcontriqsmallerdemifin} which leads to
\begin{eqnarray}
\overline{ Y_{q<1/2} } 
\simeq  E\left( z_1^q  \right) S_d
 V^{2q} \overline{ \vert  u_{\vec r} \vert^{2q} } \frac{L^{d(1-2q)}}{d(1-2q)}
\label{totqsmallerdemi}
\end{eqnarray}
The disorder-averaged exponents $\tau_{av}(q)$ 
defined in Eq. \ref{typetav} thus read
\begin{eqnarray}
 \tau_{av}(q<1/2) =  d  (2q-1) =\tau_{typ}(q<1/2)
\label{restauavqsmallerdemi}
\end{eqnarray}
and coincide the the typical exponents $\tau_{typ}(q)$,
in agreement with the 'strong multifractality' limit of Eq. \ref{sm}.

\subsection{ Disorder-averaged I.P.R. for $q>1/2$ }

For $q>1/2$,  we add the two contributions of order $\ln L$
obtained in Eqs \ref{firstcontri} and \ref{secondcontriqbiggererdemifin}.
Using $\Gamma(z+1)=z \Gamma(z)$, this leads to
\begin{eqnarray}
\overline{ Y_{q>1/2} }
&& =  1 
+ \left[  \frac{\Gamma \left( q-\frac{1}{2}\right)}{\Gamma(q)}
- 2 \frac{\Gamma \left( q+\frac{1}{2} \right)}{\Gamma(q)}
    \right]
\frac{c \sqrt{\pi} S_d V }{2  W }  
 \overline{ \vert u_{\vec r} \vert}  \ln L \nonumber \\
&& =  1 
-   \frac{\Gamma \left( q-\frac{1}{2}\right)}{\Gamma(q-1)}
\frac{ c \sqrt{\pi} S_d V }{ W }  
 \overline{ \vert u_{\vec r} \vert}  \ln L
\label{totqbiggerdemi}
\end{eqnarray}
The disorder-averaged exponents $\tau_{av}(q)$ 
defined in Eq. \ref{typetav} can be thus identified by the following expansion
\begin{eqnarray}
\overline{ Y_q(L) } \sim L^{-\tau_{av}(q)} = 1- \tau_{av}(q) \ln L
\label{DVYq}
\end{eqnarray}
This yields
\begin{eqnarray}
 \tau_{av}(q> \frac{1}{2}) = \frac{\Gamma \left( q-\frac{1}{2}\right)}{\Gamma(q-1)}
\frac{ c \sqrt{\pi} S_d V }{ W }  
 \overline{ \vert u_{\vec r} \vert}
\label{DVYqbiggerdemi}
\end{eqnarray}
This expression coincides with Eq. \ref{taulevitovrg}
obtained previously via Levitov renormalization \cite{mirlin_evers}
for the PRBM model with the correspondence $b=V/W$
if one uses  $c= {\sqrt {\frac{2}{\pi}}}$ (Eq. \ref{epsPRBM}) and
\begin{eqnarray}
\overline{ \vert  u \vert } 
= \int_{-\infty}^{+\infty}
 \frac{du}{\sqrt{2 \pi} } e^{- \frac{u^2}{2} } \vert  u \vert 
= {\sqrt {\frac{2}{\pi}}}
\label{absuGauss}
\end{eqnarray}
with $d=1$ and $S_1=2$ (to take into account the ring geometry).

\section{ Summary of the results and discussion }

\label{discussion}

In this section, we summarize and discuss the results obtained for 
the typical and the averaged multifractal spectra starting from
the perturbation formula of Eq. \ref{yq1self}.

\subsection{ Typical and averaged multifractal
 spectra $\tau_{typ}(q)$ and $\tau_{av}(q)$ }

In the region $q<1/2$, we have found that 
the typical and the averaged multifractal spectra coincide
(Eqs \ref{restautypqsmallerdemi} and \ref{restauavqsmallerdemi})
and are given by the 'strong multifractality' limit of Eq. \ref{sm}
\begin{eqnarray}
\tau_{typ}(q<1/2)= \tau_{av}(q<1/2) =  d  (2q-1) 
\label{restauqtypavqsmallerdemi}
\end{eqnarray}

In the region $q>1/2$, we have found that 
the typical and the averaged multifractal spectra are different
(Eqs \ref{restauqtypreslevy} and \ref {DVYqbiggerdemi})
\begin{eqnarray}
\tau_{typ}(q>1/2) && = S_d 
\frac{c V }{W} \overline{\vert u_{\vec r} \vert}  \pi 
\left[ q - \frac{ 1 }{ \sin ( \frac{\pi}{2q} )  }    \right] \nonumber \\
 \tau_{av}(q> 1/2) && = S_d \frac{ c  V }{ W }  
 \overline{ \vert u_{\vec r} \vert} \sqrt{\pi} 
\frac{\Gamma \left( q-\frac{1}{2}\right)}{\Gamma(q-1)}
\label{restauqtypavqbiggerdemi}
\end{eqnarray}
where the results for $ \tau_{av}(q> 1/2)$ coincides
with the result of Eq. \ref{taulevitovrg}
obtained previously via Levitov renormalization \cite{mirlin_evers}.

Our conclusion is thus that the critical value $q_c$
where the typical 
and averaged spectra separate (Eq. \ref{criterionbetaq}) is 
\begin{eqnarray}
q_c=\frac{1}{2}
\label{qcdemi}
\end{eqnarray}
in contrast with the other value $q_c \simeq 2.4$ predicted in \cite{mirlin_evers},
on the basis of the vanishing of the averaged singularity spectrum $f_{av}(\alpha)$
(see below Eq. \ref{a+q+av}).
Moreover,  we expect that the distribution $P_q(y_q)$
of the rescaled variable $y_q=Y_q(L)/Y_q^{typ}(L)$ 
will decay in the scaling regime with the power law of Eq. \ref{defbetaq}
of exponent $(1+\beta_q)$ that will coincide with the exponent $(1+\mu_q)$
describing the distribution of $\Sigma_q$ (see Eq. \ref{lawsigmaq})
\begin{eqnarray}
\beta_q= \mu_q = \frac{1}{2q} = \frac{q_c}{q}
\label{resbetaq}
\end{eqnarray}
in agreement with Eq. \ref{betaqafterqc}.

\subsection{ Consequences for the typical and averaged singularity spectra
  $f_{typ}(\alpha)$ and $f_{av}(\alpha)$ }

Besides the multifractal exponents $\tau_{typ}(q)$ and $\tau_{av}(q)$,
it is usual to introduce the typical and averaged singularity spectra
 $f_{typ}(\alpha)$ and $f_{av}(\alpha)$ which describe 
the numbers ${\cal N}_L^{typ,av}(\alpha)$
of points $\vec r$ in a sample of size $L^d$, 
where the weight $\vert \psi_L(\vec r)\vert^2$
scales as $L^{-\alpha}$ \cite{mirlinrevue}
\begin{eqnarray}
{\cal N}_L^{typ,av}(\alpha) \oppropto_{L \to \infty} L^{f_{typ,av}(\alpha)}
\label{nlalpha}
\end{eqnarray}
The I.P.R. $Y_q$ can be then rewritten
as integrals over $\alpha$
\begin{equation}
Y_q^{typ,av}(L)  
\simeq \int d\alpha \ L^{f_{typ,av}(\alpha)} 
\ L^{- q \alpha} \opsimeq_{L \to \infty} L^{ - \tau_{typ,av}(q) }
\label{ipr}
\end{equation}
 The exponents $\tau_{typ,av}(q)$ 
can be obtained via a saddle-point
calculation in $\alpha$ to obtain the Legendre
transform formula \cite{janssenrevue,mirlinrevue}
\begin{eqnarray}
 -\tau_{typ,av}(q)  =  \opmax_{\alpha} \left[ f_{typ,av}(\alpha) - q \alpha  \right]
\label{legendre}
\end{eqnarray}

At leading order, the 'strong multifractality' limit of Eq. \ref{sm}
( or Eq. \ref{restauqtypavqsmallerdemi} above)
corresponds to the singularity spectra \cite{mirlin_evers}
\begin{eqnarray}
f_{typ}(\alpha)=f_{av}(\alpha)= \frac{\alpha}{2} \ \ {\rm for} \ \ 0\leq \alpha \leq 2d
\label{ftypavsm}
\end{eqnarray}
The typical exponent $\alpha_{typ}$ where $f_{typ}(\alpha_{typ})=d$
thus corresponds to the maximal value $\alpha_{typ}=\alpha_{max}=2d$.
The singularity spectrum vanishes only at the other boundary $\alpha_{min}=0$.
From the point of view of the Legendre transform of Eq. \ref{legendre},
this case is singular since the saddle $\alpha^*(q)$
is concentrated on the two values $\alpha^*(q<1/2))=\alpha_{max}=2d$
and $\alpha^*(q>1/2))=\alpha_{min}=0$.

\begin{figure}[htbp]
 \includegraphics[height=5cm]{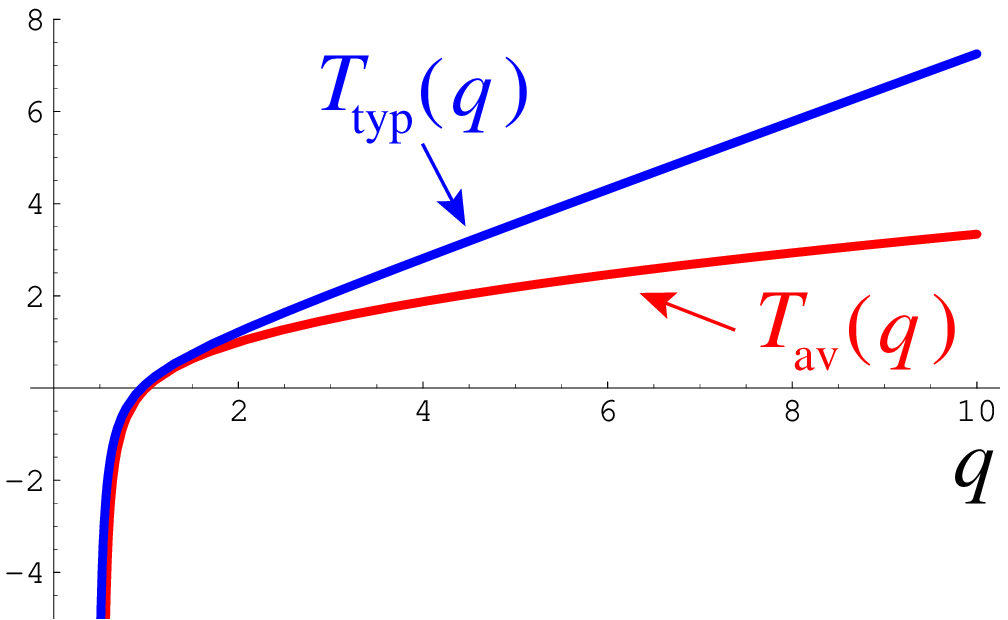}
 \hspace{1cm}
 \includegraphics[height=5cm]{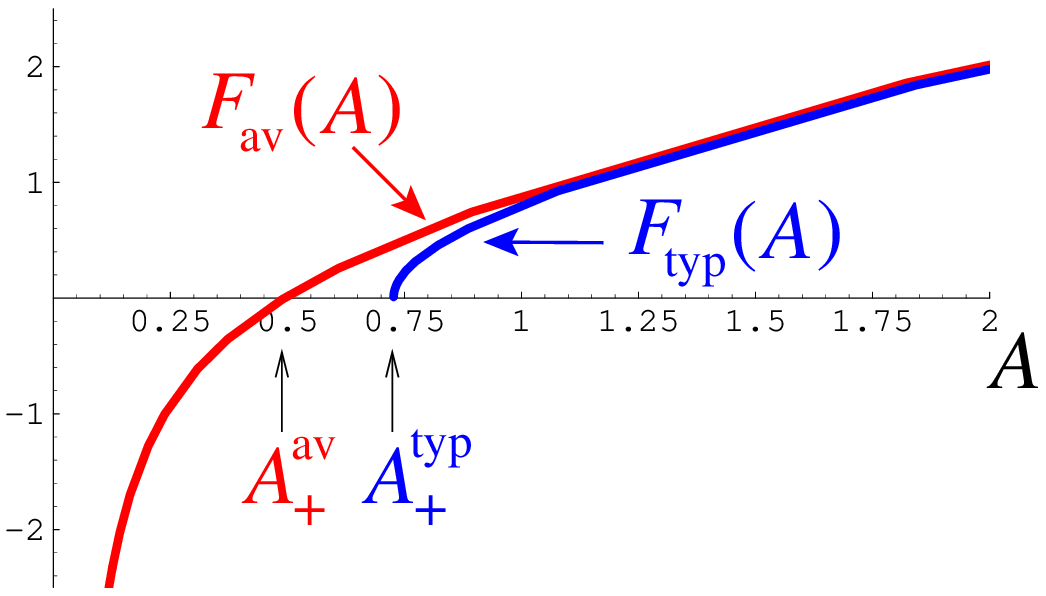}
\vspace{1cm}
\caption{ (Color on line) 
(a) Comparison of the typical and averaged multifractal spectra 
 $T_{typ}(q>1/2)$ and $T_{av}(q>1/2)$ of Eq. \ref{tavtyp} :
they are close near $q \to (1/2)^+$ (Eq. \ref{tavtypsing})
but are very different at large $q$ (Eq. \ref{tavtypinfty}).
(b) Comparison of the corresponding typical and averaged
singularity spectra $F_{typ}(A)$ and $F_{av}(A)$ :
$F_{typ}(A)$ exactly terminates at the point $
A_+^{typ}$ of Eq. \ref{a+typ} corresponding to $q=+\infty$, 
where it vanishes $F_{typ}(A_+^{typ})=0$, whereas $F_{av}(A)$ vanishes
at another value $A_+^{av}$ corresponding to $q_+^{av}$ (see Eq. \ref{a+q+av})
and continues in the negative domain $F_{av}(A)<0$.}
\label{figmultif}
\end{figure}

Let us now take into account the small corrections of Eq. \ref{restauqtypavqbiggerdemi} in the domain $q>1/2$, where we factorize the small prefactor
$\epsilon=S_d 
\frac{c V }{W} \overline{\vert u_{\vec r} \vert} \frac{\pi}{2}$
(this choice of constants in $\epsilon$ has been preferred
to have the same normalization of $T_{av}(q)$ as in \cite{mirlin_evers})
\begin{eqnarray}
T_{typ}(q>1/2) \equiv \frac{\tau_{typ}(q>1/2)}{\epsilon} && =
2 \left[ q - \frac{ 1 }{ \sin ( \frac{\pi}{2q} )  }    \right] \nonumber \\
T_{av}(q>1/2) \equiv \frac{\tau_{av}(q>1/2)}{\epsilon} && =
\frac{2 \Gamma \left( q-\frac{1}{2}\right)}{ \sqrt{\pi} \Gamma(q-1)}
\label{tavtyp}
\end{eqnarray}
The two multifractal spectra $T_{typ}(q>1/2)$ and $T_{av}(q>1/2)$ are shown on Fig. \ref{figmultif} (a) for comparison : there are very close in the region $q \to 1/2$
where they present the same singularity
\begin{eqnarray}
T_{typ}(q>1/2) && \opsimeq_{q \to (1/2)^+ } - \frac{1}{\pi \left( q-\frac{1}{2} \right)} \nonumber \\
T_{av}(q>1/2)  && \opsimeq_{q \to (1/2)^+ } - \frac{1}{\pi \left( q-\frac{1}{2} \right)}
\label{tavtypsing}
\end{eqnarray}
but are completely different at large $q$ with the following asymptotic behaviors
\begin{eqnarray}
T_{typ}(q>1/2) && \opsimeq_{q \to +\infty } 
2 \left( 1 - \frac{ 2 }{ \pi  }    \right) q  \nonumber \\
T_{av}(q>1/2)  && \opsimeq_{q \to +\infty } \frac{2 }{ \sqrt{\pi}} q^{1/2}
\label{tavtypinfty}
\end{eqnarray}

As discussed in \cite{mirlin_evers}, the disorder-averaged singularity spectrum
$f_{av}(\alpha)$ takes the scaling form
\begin{eqnarray}
f_{av}(\alpha)= \epsilon F_{av} \left( A= \frac{\alpha}{\epsilon} \right)
\label{fav}
\end{eqnarray}
where $F_{av}(A)$ is the Legendre transform of $T_{av}(q)$ :
its properties have been described in \cite{mirlin_evers}.
In particular, it vanishes at $A_+^{av}$ corresponding to $q_+^{av}$ with the numerical values \cite{mirlin_evers}
\begin{eqnarray}
A_+^{av} && \simeq 0.51 \nonumber \\
q_+^{av} &&\simeq 2.405
\label{a+q+av}
\end{eqnarray}

Similarly, the typical singularity spectrum $f_{typ}(\alpha)$ takes the scaling form  \begin{eqnarray}
f_{typ}(\alpha)= \epsilon F_{typ} \left( A= \frac{\alpha}{\epsilon} \right)
\label{ftyp}
\end{eqnarray}
where $F_{typ}(A)$ is the Legendre transform of $T_{typ}(q)$.
From the asymptotic linear behavior of $T_{typ}(q)$ (see Eq. \ref{tavtypinfty}),
one obtains that $F_{typ}(A)$ exactly terminates at the point
 \begin{eqnarray}
A_+^{typ} = 2 \left( 1 - \frac{ 2 }{ \pi  }    \right) \simeq 0.727
\label{a+typ}
\end{eqnarray}
corresponding to $q=+\infty$, where it vanishes $F_{typ}(A_+^{typ})=0$.
The two singularity spectra $F_{av}(A)$ and $F_{typ}(A)$ are shown
 in Fig. \ref{figmultif} (b) for comparison.

The fact that $F_{typ}(A)$ exists only in the region where it remains positive
$F_{typ}(A) \geq 0 $ is a standard property of typical spectrum 
\cite{mirlinrevue}. What is surprising however is that
 the typical and averaged singularity spectra differ even in the region
where there are positive, whereas the standard picture is that
$F_{typ}(A)=F_{av}(A) \theta( F_{av}(A) >0)$ \cite{mirlinrevue}.
 Equivalently, in this standard picture \cite{mirlinrevue}, 
the typical spectrum is expected to be exactly linear $T_{typ}(q) = A_+ q$
for $q>q_+$ meaning that in the Legendre calculation of Eq. \ref{legendre},
the saddle value remains frozen at $A_{typ}(q \geq q_+)=A_+$.
Here we have found instead that the typical spectrum $T_{typ}(q)$
is not exactly linear
in $q$ in the region where it is different from $T_{av}(q)$,
and that the saddle value $A_{typ}(q)$ of the Legendre transform of Eq. \ref{legendre} reaches the termination point $A_+^{typ}$ only in the asymptotic regime $q \to +\infty$. 

\subsection{ Discussion on the method }

The perturbative calculation of the multifractal spectrum in the strong multifractality regime can only be very singular since one starts from a complete
localized basis to construct multifractal
 critical eigenvectors via perturbation.
To face this difficulty, two strategies have been proposed :

(i)  The powerful Levitov renormalization method \cite{levitov,mirlin_evers,fyodorov} performs iterative changes of bases to take into account the resonances that occur at various scales. This approach has been reformulated as
some type of 'virial expansion' in Refs \cite{oleg1,oleg2,oleg3,oleg4}.

(ii) In the present paper, we have proposed instead
to use the standard perturbation theory of quantum mechanics.
It is thus simpler than (i), since we work in the initial completely localized basis. However, since the random perturbation terms are singular, the 
essential point in this approach is that the I.P.R. should be computed with
Eq. \ref{yq1self}, where the perturbation terms of the eigenvectors appear
both in the numerator and in the denominator : this ratio is then regular,
because any potential divergence appearing in the numerator is compensated
by the corresponding divergence in the denominator.
In the present paper, we have described in detail 
how the first-order expression of the
perturbed eigenvector allows to obtain the leading order of the multifractal
spectrum in various regions of $q$.
A natural question is how this approach can be pursued at higher orders.
We stress that one should not use the standard Rayleigh-Schr\"odinger
expressions for the normalized perturbed eigenvector
(since these expressions are in fact based on the perturbative expansion of the
normalization, which is singular here as explained above). 
We believe that the correct formulation of our approach at higher orders
involve the perturbative expansion of the eigenvector
\begin{eqnarray}
\vert \phi_{\vec r}^{tot} > && = \vert \phi_{\vec r}^{(0)} > + 
\sum_{n=1}^{+\infty}   \vert \phi_{\vec r}^{(n)} >
= \vert \vec r >+\sum_{n=1}^{+\infty}   \vert \phi_{\vec r}^{(n)} >
\label{phitot}
\end{eqnarray}
in the so-called intermediate normalization defined by 
\begin{eqnarray}
< \phi_{\vec r}^{(0)}\vert \phi_{\vec r}^{tot} > = < \vec r \vert \phi_{\vec r}^{tot} > = 1
\label{intermediate}
\end{eqnarray}
so that all corrections are orthogonal to the zeroth order term
$\vert \phi_{\vec r}^{(0)} >=\vert \vec r >$
\begin{eqnarray}
< \phi_{\vec r}^{(0)}\vert \phi_{\vec r}^{(n)} > 
= < \vec r \vert \phi_{\vec r}^{(n)} >=0 \ \ {\rm for } \ \ n \geq 1
\label{orthog}
\end{eqnarray}
The corresponding I.P.R. of Eq. \ref{defipr} should be then obtained as
\begin{eqnarray}
Y_q = \frac{ \displaystyle \sum_{\vec r \ '}
 \vert \phi_{\vec r}^{tot}(\vec r \ ') \vert^{2q} }
{ \left[  \displaystyle \sum_{\vec r \ '}  \vert \phi_{\vec r}^{tot}(\vec r \ ') \vert^{2}  \right]^q }
= \frac{1+ \displaystyle \sum_{\vec r \ ' \ne \vec r}
 \vert \phi_{\vec r}^{tot}(\vec r \ ') \vert^{2q} }
{ \left(1+ \displaystyle\sum_{\vec r \ '\ne \vec r} 
\vert  \phi_{\vec r}^{tot}(\vec r \ ') \vert^2 \right)^q }
  \label{yq}
\end{eqnarray}

\section{ Conclusions and perspectives }

\label{conclusion}

In summary, we have show that that the strong multifractality
regime of Anderson tight-binding models in dimension $d$
with critical long-ranged hoppings
 can be studied via the standard perturbation theory for eigenvectors
in quantum mechanics. The Inverse Participation Ratios $Y_q(L)$,
which are the order parameters of Anderson transitions, then involve
 weighted L\'evy sums of broadly distributed variables,
as a consequence of the presence of on-site random energies in the denominators
of the perturbation theory.
We have computed at leading order
 the typical and disorder-averaged multifractal spectra
$\tau_{typ}(q)$ and $\tau_{av}(q)$ as a function of $q$.
For $q<1/2$, we have found the non-vanishing
limiting spectrum $\tau_{typ}(q)=\tau_{av}(q)=d(2q-1)$ as $V \to 0^+$,
that had been obtained previously in \cite{mirlin_evers}
only indirectly via the symmetry relation of Eq. \ref{symtauq}.
For $q>1/2$, we have obtained the same result for 
disorder-averaged spectrum $\tau_{av}(q)$ at order $O(V)$
as obtained previously
via the Levitov renormalization method \cite{mirlin_evers}.
This agreement between these two completely different approaches
is a good indication in favor of the exactness of this result.
But in addition, our present approach allows to 
compute explicitly the typical spectrum 
(that has not been computed via Levitov renormalization) 
: it is also of order $O(V)$
but has a different $q$-dependence $\tau_{typ}(q) \ne \tau_{av}(q)$ for $q>q_c=1/2$,
and is not exactly linear in this regime, in contrast with the standard picture
\cite{mirlinrevue}.
As a consequence, we have found that the corresponding singularity spectra 
$f_{typ}(\alpha)$ and $f_{av}(\alpha)$ differ even in the positive region $f>0$,
in contrast with the standard picture where they coincide in the positive region
$f_{typ}(\alpha)=f_{av}(\alpha) \theta( f_{av}(\alpha) >0)$ \cite{mirlinrevue},
and that the saddle value $A_{typ}(q)$ of the Legendre transform reaches the termination point $A_+^{typ}$ where $f_{typ}(A_+^{typ} )=0 $
only in the limit $q \to +\infty$. 

In conclusion, the present work based on a pedestrian perturbative explicit approach
thus questioned important statements of the standard picture
of multifractality at Anderson transitions. 
We hope that it will stimulate further studies to better understand
the differences between typical and averaged multifractal spectra.

\section*{ Acknowledgements }

It is a pleasure to thank Y. Fyodorov for an interesting discussion
and for mentioning to us Refs \cite{oleg1} and \cite{oleg2}.

\end{document}